\documentclass[runningheads]{llncs}

\usepackage{graphicx}

\usepackage[utf8]{inputenc} %
\usepackage[T1]{fontenc}    %
\usepackage{hyperref}       %
\usepackage{url}            %
\usepackage{booktabs}       %
\usepackage{amsfonts}       %
\usepackage{nicefrac}       %
\usepackage{microtype}      %
\usepackage{amsmath}
\usepackage{lipsum}
\newcommand{\ra}[1]{\renewcommand{\arraystretch}{#1}}
\usepackage{caption}
\usepackage{float}
\usepackage{subcaption}

\makeatletter
\newcommand{\printfnsymbol}[1]{%
  \textsuperscript{\@fnsymbol{#1}}%
}
\makeatother

\usepackage{color}

\begin{document}
\title{Semantic Segmentation of Histopathological Slides for the Classification of Cutaneous Lymphoma and Eczema}
\titlerunning{Semantic Segmentation of Histopathological Slides}

\author{Jérémy Scheurer\inst{1}\thanks{Equal Contribution}%
\and Claudio Ferrari\inst{1}\printfnsymbol{1}%
\and Luis {Berenguer Todo Bom}\inst{1}\printfnsymbol{1}%
\and Michaela Beer\inst{2}%
\and Werner Kempf \inst{2}%
\and Luis Haug\inst{1}%
} 

\authorrunning{Scheurer et al.}

\institute{
ETH Zürich, Zürich, Switzerland\\
\email{\{ferraric,jeremys,beluis,lhaug\}@ethz.ch} \and
Kempf und Pfaltz Histologische Diagnostik, Zürich, %
\email{\{Michaela.Beer,Werner.Kempf\}@kempf-pfaltz.ch}}

\maketitle              %

\begin{abstract}
Mycosis fungoides (MF) is a rare, potentially life threatening skin disease, which in early stages clinically and histologically strongly resembles Eczema, a very common and benign skin condition. In order to increase the survival rate, one needs to provide the appropriate treatment early on. To this end, one crucial step for specialists is the evaluation of histopathological slides (glass slides), or Whole Slide Images (WSI), of the patients' skin tissue. We introduce a deep learning aided diagnostics tool that brings a two-fold value to the decision process of pathologists. First, our algorithm accurately segments WSI into regions that are relevant for an accurate diagnosis, achieving a Mean-IoU of $69\%$ and a Matthews Correlation score of $83\%$ on a novel dataset. Additionally, we also show that our model is competitive with the state of the art on a reference dataset.
Second, using the segmentation map and the original image, we are able to predict if a patient has MF or Eczema. We created two models that can be applied in different stages of the diagnostic pipeline, potentially eliminating life-threatening mistakes. The classification outcome is considerably more interpretable than using only the WSI as the input, since it is also based on the segmentation map. Our segmentation model, which we call EU-Net, extends a classical U-Net with an EfficientNet-B7 encoder which was pre-trained on the Imagenet dataset.

\end{abstract}

\section{Introduction}
Mycosis fungoides is a slowly progressing but potentially life-threatening neoplastic skin disease derived from lymphocytes. It is the most common form of cutaneous T-cell lymphoma and starts with erythematous patches that grow and develop to plaques and tumors in later disease stages. The tumor cells can then metastasize to other organs, leading to the patient's death. If, however, appropriate treatment is applied and progression can be stopped or controlled, the patients encounter an indolent disease course and an almost normal survival rate. Eczema, on the other hand, is a very common and benign inflammatory skin disease which strongly resembles the clinical and histopathological features of MF in its early stages \cite{rovner2015influence}. The distinction between early MF and Eczema is of utmost importance to apply an appropriate treatment and prevent a development of MF to later stages \cite{santucci2000efficacy}.

The differentiation between the two diseases can however be very difficult due to overlapping histopathological features \cite{PIMPINELLI20051053}, \cite{KEMPF2015655}.  This is also apparent in the inter-rater variability, i.e., the agreement between pathologists, of only 48 percent \cite{guitart} when diagnosing patients with MF. Another study with pathologists specialized on cutaneous lymphomas, also showed a misclassification error rate of $21.51\%$ and a inter-rater variability with a Cohen’s kappa value that ranged from $0.3762$ to $0.4332$, which means the pathologists only had a fair to moderate agreement \cite{PIMPINELLI20051053}. These numbers paint a grim picture, especially as there are high risks involved in diagnosing MF. Not recognizing it early on can lead to delayed diagnosis and treatment. On the other hand, over-diagnosing benign lesions could lead to unnecessary and potentially harmful therapy. Furthermore, MF is considered to be a rare disease with an incidence rate of $4$ in $100'000$ patients \cite{garbe2013dermatologische} which also leads to a lack of specialized experts, especially in the field of cutaneous lymphomas. A computer assisted diagnosis system based on deep learning algorithms to detect MF could help to reduce the workload of pathologists and furthermore reduce misinterpretations by non-experts.

\begin{figure}
    \centering
    \includegraphics[width=\textwidth]{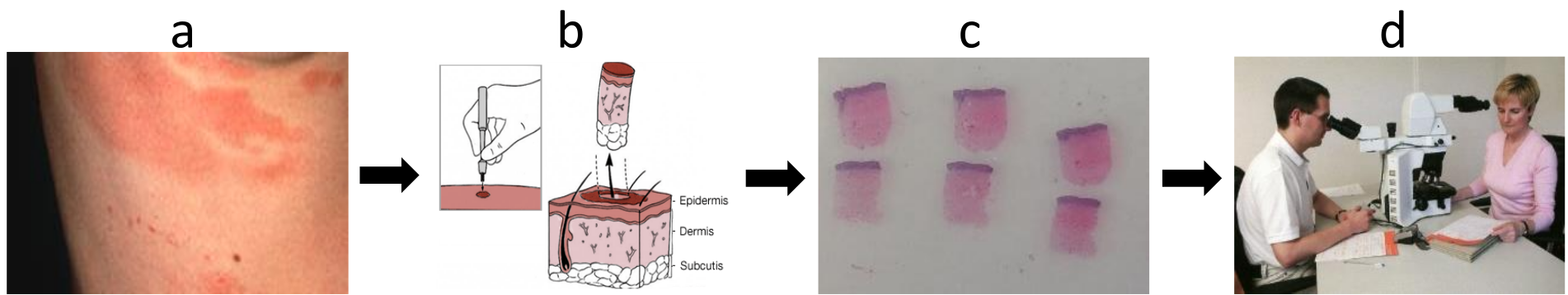}
    \caption{ The process starts from the medical examination (a), then a biopsy is performed (b), and sent to a histopathology laboratory where it is processed, stained on a glass slide (c), and subsequently examined under a microscope by one or more experienced pathologists (d).}
    \label{fig:medical_process}
\end{figure}

In order to diagnose a patient with MF, or the absence of it, pathologists look at glass slides or WSI of scanned histopathological skin biopsy specimens (see Figure \ref{fig:medical_process}). We introduce a novel dataset of WSI and annotations from the pathology laboratory Kempf und Pfaltz Histologische Diagnostik\footnote{\url{kempf-pfaltz.ch}}. The annotations were created by pathologists and specially trained biologists and contain pixel-wise class annotations for the relevant categories: spongiosis, epidermis and ``rest''. The epidermis is the outmost layer of the skin, and spongiosis are regions inside the epidermis where there has been an abnormal accumulation of intercellular fluid. The class "rest" denotes other tissue (meaning tissue that is neither epidermis nor spongiosis) or image background. Apart from the number and distribution of neoplastic lymphocytes within the epidermis as one of the diagnostic criteria, the presence or absence of spongiosis in certain areas of the epidermis is one among other essential factors for pathologists to make a correct diagnosis \cite{PIMPINELLI20051053}. To this end, we trained a deep learning model on these images and learned to predict a pixel-wise segmentation. This, in itself, is already very useful, as it is the basis for making a right classification and the process of manually annotating a WSI is very tedious and time consuming. Additionally, all the WSI are labeled as manifesting a case of MF or Eczema. With this data, we train another model with which we attempt to output a correct diagnosis, i.e., if a patient has MF or Eczema, using only the WSI and its corresponding segmentation map as input.

In this paper, we introduce a U-Net \cite{DBLP:journals/corr/RonnebergerFB15} architecture with an EfficientNet \cite{tan2019efficientnet} encoder which we call EU-Net \cite{Yakubovskiy:2019}. We argue that by using the EU-Net that was pre-trained on the ImageNet  dataset \cite{deng2009imagenet}, we get very accurate segmentation predictions on the dataset from Kempf und Pfaltz Histologische Diagnostik. To validate our approach against a baseline, we further show that it yields competitive results on the test set of \cite{Oskal2019}. In addition, we believe that our classification predictions on the former dataset could be helpful to pathologists as a diagnostic aid tool, for the process of screening cases. Our contributions are two-fold: 
\begin{enumerate}
    \item We use an EU-Net, pre-trained on the ImageNet dataset \cite{deng2009imagenet}, to  train and test on our own dataset, which provides a challenging setting for semantic segmentation, due to having annotations for pixel-classes epidermis, spongiosis and ``rest''. Furthermore, we achieve competitive results on a benchmark test set of \cite{Oskal2019}, which has only annotations for the classes epidermis and ``rest''.
    \item We use the segmentation map generated by our architecture as an additional feature to the WSI, to classify them into classes MF or Eczema. We show that the resulting models could be a useful diagnostic aid tool.
\end{enumerate}

\section{Related Work}
Although there have been many related publications made on the problem of semantic segmentation of epidermis, as far as we know, there have been no attempts at segmentation of spongiosis and epidermis as is our case. However, much work has been done on similar and related tasks, from which we drew inspiration. 

Traditional computer vision approaches have been used for semantic segmentation of epidermis. Heggerty at al.~\cite{haggerty2014segmentation} created an algorithm that combines enhanced color information with general image intensity information. The segmentation results from thresholding, morphological processing and object classification rules. Lu et al.~\cite{lu2015automated} utilize a simpler method, mainly consisting of thresholding the red channel, followed by a shape analysis. This is done on a down-sampled image, and then a template matching method is used to obtain the segmentation on the original image. Xu et al.~\cite{hongming2015epidermis} utilize an approach similar to \cite{lu2015automated}, followed by the application of the $k$-means algorithm for a more fine-grained segmentation. This approach is also used in \cite{xu2018automated} where they propose a multi-resolution framework. Kłeczek et al. \cite{kleczek2017automated} perform segmentation based on the information about shape, distribution of transparent regions and distribution and concentration of hematoxylin and eosin stains.

There have been many medical imaging applications where deep learning was able to improve on the state of the art \cite{anthimopoulos2018semantic}  \cite{DBLP:journals/corr/RonnebergerFB15}. Transfer learning has also been applied to several medical imaging challenges such as %
\cite{christodoulidis2016multisource}, 
\cite{wang2016deep}.
There have been many cases where deep learning was applied to histopathological slides, such as %
\cite{aresta2019bach}, \cite{chan2019histosegnet}, \cite{Fu2018SegmentationOH},  
\cite{xie2019interpretable}
 and a more comprehensive overview can be gained through \cite{inproceedingsdabass}.
 To our knowledge, there has only been one attempt at semantic segmentation of epidermis using a deep learning approach \cite{Oskal2019}, specifically using an adapted U-Net~\cite{DBLP:journals/corr/RonnebergerFB15}. Their modifications to the original architecture include the following: halving the number of feature channels, using zero-padded convolutions, a different activation function, batch normalization %
 and dropout. %
 Finally, they also use image post-processing methods to improve their predictions. They trained and evaluated their model on a combination of datasets from the University of Michigan\footnote{\url{https://www.pathology.med.umich.edu/slides/search.php?collection=Andea\&dxview=show}, Accessed Nov 2019} and Columbia\footnote{\url{http://histo.anat.ubc.ca/PATHOLOGY/Anatomical\%20Pathology/DermPath/}, Accessed Nov 2019}. This combined dataset, which we will refer to as the Michigan-Columbia dataset, contains histopathological slides with labeled epidermis. Lastly, \cite{kwok2018multiclass} applies deep learning on WSI for breast-cancer classification. This paper was the main inspiration for our attempt at binary classification of MF vs.\ Eczema.

\section{Data}
\subsection{Segmentation Dataset} \label{section:Seg_data}
We use a novel dataset created by Kempf und Pfaltz Histologische Diagnostik. This dataset, which we will refer to as the MF/E-Segmentation dataset, consists of $164$ high resolution annotated Hematoxylin-Eosin stained WSI (see Figure~\ref{fig:mf_e_dataset_segmentation_predictions}). Out of these, $60$ slides are labeled as MF and $104$ slides as Eczema (E). Note that for certain patients we have multiple images and annotations. However, when splitting our data into train and test sets, we made sure to put slides of the same patient into the same split.  

These slides were divided into training set ($60\%$), validation set ($20\%$) and test set ($20\%$). Since we have an imbalanced dataset, we maintained the same imbalance ratio in each split. On the training set we employed a patch extraction method similar to \cite{Oskal2019}, motivated by our large class imbalance and the fact that the original WSI would not fit on a standard GPU. From the train slides, we extracted $78'125$ patches with a resolution of $512 \times 512$. The slides on the validation and test set were simply cut into patches with a resolution of $512 \times 512$ for the validation and $4096 \times 4096$ for the test set. Taking larger patches for the test set was done to be consistent with our results on the Michigan-Columbia dataset.

\subsection{Patch Extraction Technique} \label{patching_tech}

 Our patch extraction method is based on the idea from \cite{Oskal2019} to solve class imbalance, but adapted to work in a multi-class setting. In broad terms, this technique aims to under-sample the majority classes, and over-sample the minority classes. This is done by defining a patch as belonging to a specific type if it contains a high-enough percentage of pixels of said type. The types of patches are the three classes epidermis, spongiosis and ``rest''. However, the class ``rest'' can consist of pixels of either ``other tissue'' (i.e. non-epidermis and non-spongiosis tissue) or image background (usually white pixels). Therefore, we decided to split the class ``rest'' into two subclasses, to ensure we have both ``other tissue'' and background pixels labelled as class ``rest'' in our training set. The goal is to end up with a similar number of patches per type for each of the slides.

The patch extraction process works as follows: First, we randomly pick a pixel from the WSI, which will be the top-left corner of the patch. Then, we ensure that there is a distance of at least $100$ pixels from any previously picked top-left corner. The next step is determining which of the four types of patch this will be counted towards. The conditions are the following:

\begin{itemize}
  \item Background patch, if all pixels are white
  \item Spongiosis patch, if $>20\%$ of pixels are spongiosis
  \item Epidermis patch, if $>40\%$ of pixels are epidermis or spongiosis
  \item Other tissue patch, if it did not fall into any previous category
\end{itemize}

The threshold for epidermis was chosen according to \cite{Oskal2019}, while the threshold for spongiosis was chosen lower since this helped finding enough patches for spongiosis, which is the rarest class.
We repeated this process until we had enough patches for each type. We were able to extract between $19'000$ and $20'000$ patches for each type, totaling $78'125$ patches. More precisely, we have $19'118$ spongiosis patches, $19'807$ epidermis patches, $19'600$ background patches and $19'600$ other tissue patches.

\subsection{Binary Classification Dataset}
For the classification into the classes Eczema and MF, we created a dataset of 307 WSI (209 of class Eczema and 98 of class MF), originating from 93 different patients. We will henceforth refer to this dataset as MF/E-Classification. The reason that this dataset contains more WSI than the MF/E-Segmentation dataset, is because we do not actually need an annotated segmentation label for each WSI. This allows us to extract more WSI per patient. However, even though we have more WSI, we still have much less data than for the segmentation. The reason is, because we can no longer extract patches from a WSI. This is due to the fact that we do not know what parts of the slide will contribute to the binary classification decision. We only have the labels at the WSI level. In addition, we have a class imbalance of around 2 : 1 of Eczema to MF, which makes the task even harder. Finally, we resized each WSI to 4096$\times$4096 in order to fit them on a GPU.

\subsection{Michigan-Columbia Dataset} \label{michigancolumbiadataset}

To the best of our knowledge, there seems to have been no research in semantic segmentation of histopathological slides for the classification of cutaneous lymphoma and Eczema. Specifically, there is no previous work conducted on the segmentation of WSI into the pixel-level classes epidermis, spongiosis and ``rest''. This inhibits our ability to benchmark our proposed models against alternative solutions on the exact same task. However, we were able to compare our approach with a U-Net approach by \cite{Oskal2019} on a segmentation task of WSI into the pixel-level classes epidermis and ``rest''. For this reason, we contacted the authors of \cite{Oskal2019} and requested access to their dataset, which is based on the datasets of Michigan and Columbia University and contains a total of 69 WSI. Our train and test set used the same split of WSI as \cite{Oskal2019}. However, it is important to note that they did not share their actual train set with us, but only the patch extraction script to generate it from the WSI. Due to the stochasticity of the patch extraction script, our actual train set is thus not identical to theirs. Note, however, that our test sets are identical, as one does not apply the patch extraction to it. Since these authors did a comprehensive comparison of various models on their dataset, we can compare their results with the performance of our developed models, thereby benchmarking our models.

\begin{figure}
    \centering
     \hfill
     \begin{minipage}{\textwidth}
        \centering
        \includegraphics[width=\textwidth]{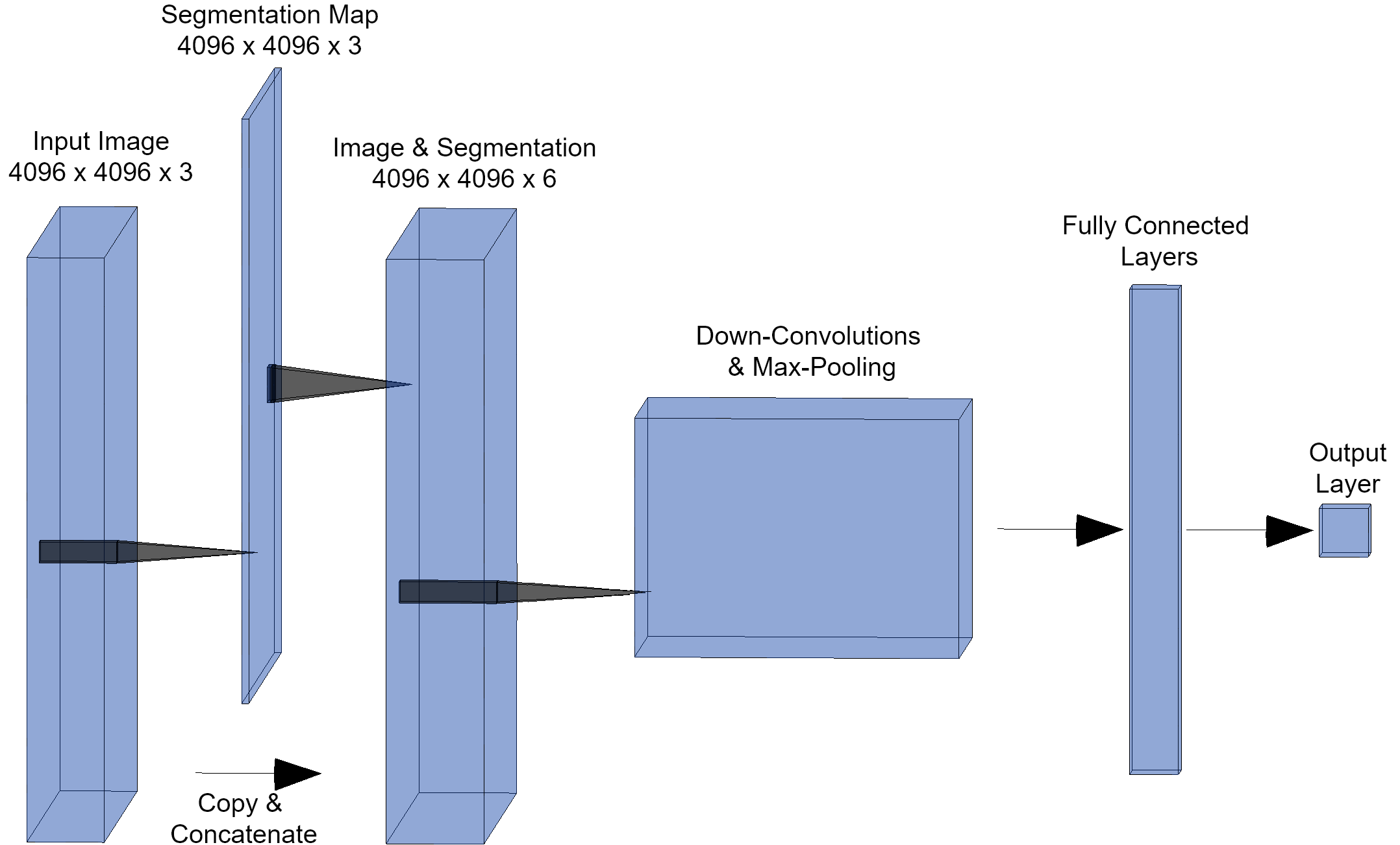}
        \captionof{figure}{Neural Network Architecture used for binary classification. Utilizes our best segmentation model to create the segmentation map and concatenate it to the original image. Note that we also evaluate the same architecture without the additional segmentation map. See the results in section \ref{results:classification}.}
        \label{fig:nn_arch}
     \end{minipage}
\end{figure}

\section{Methods} 
We introduce an EU-Net \cite{Yakubovskiy:2019} for the task of segmentation and classification of WSI. Moreover, we use a standard U-Net \cite{DBLP:journals/corr/RonnebergerFB15} in order to have a baseline on our own dataset. For the binary classification our best performing EU-Net model was used for creating the segmentation maps.

\subsection{U-Net}
Our U-Net architecture is very similar to the standard U-Net proposed in \cite{DBLP:journals/corr/RonnebergerFB15} (see Appendix~\ref{Appendix:unet} for an illustration), with a few minor exceptions which are due the specifics of our task. The standard U-Net starts with a contracting part, which exhibits the typical convolutional network architecture of stacking multiple $3\times3$ convolutions, followed by max pooling for down sampling. This is followed by a expansive part, which consists of up-convolution for up sampling.

\subsection{EU-Net}
Our proposed method is an EU-Net \cite{Yakubovskiy:2019}, which has a similar encoder-decoder structure as the basic U-Net \cite{DBLP:journals/corr/RonnebergerFB15}. It differs from the U-Net in that it has an EfficientNet-B7 \cite{tan2019efficientnet} encoder and a corresponding decoder. This results in a much larger model with around 80 million parameters. In general, creating an architecture from scratch for a specific task is a very challenging endeavor. This is why  \cite{tan2019efficientnet} use multi-objective neural architecture search to come up with a basic EfficientNet-B0 architecture that trades off accuracy and FLOP's. They further introduce a scaling method which increases the model size by scaling the width, depth and resolution of the convolutions and eventually yields the EfficientNet-B7 architecture. This model was able to achieve state of the art performance on the Imagenet dataset \cite{deng2009imagenet}, while at the same time being much smaller and faster. They further show that their architecture performs well on transfer-learning tasks. 

These properties were the reason we chose to replace the encoder structure of the U-Net with an EfficientNet-B7. Specifically, the input is encoded with the EfficientNet-B7, and the output of the last encoder-layer is then fed to the decoder. The decoder consists of 5 up-convolution modules where each module contains an up-convolution, followed by a normal convolution (see Appendix \ref{Appendix:eunet}). In the up-convolution modules we use batch normalization %
after the up-convolution and our last layer has either a sigmoid or softmax activation function, depending on whether the task is binary or multi-class classification. Lastly, the whole model was pre-trained on the ImageNet dataset \cite{deng2009imagenet}. For more details about the training procedure we refer the reader to Appendix \ref{Appendix:efficientnet}. For our model we used a library provided by \cite{Yakubovskiy:2019}, which allowed us to also experiment with different encoders and general segmentation architectures.

\subsection{Binary Classification of MF vs.\ Eczema}
Given a WSI and its predicted segmentation map, we built an architecture that extracts features from these two inputs, which we then use to do a binary classification (see Figure~\ref{fig:nn_arch}). Specifically, we used our best performing model, the EU-Net, to create the segmentation map and concatenate it with the original image. This is followed by 4 modules, where each module consists of a down-convolution with a kernel size of 3, followed by a max-pooling layer. The output of the 4 modules is then fed into two fully-connected layers with 128 and 1 output units. Note that all layers use a ReLU activation, except the last layer which uses a sigmoid. 

To investigate whether the segmentation map is actually needed as an additional input to the WSI, we do an ablation study where we train the same model with and without the segmentation map as input. The results can be found in section \ref{results:classification}. Additionally, we would like to point out that using the segmentation map for the classification provides the benefit that a pathologist can look at the actual disease prediction, and at the same time use the predicted segmentation map as a means to interpret the results.

\section{Results}

\subsection{Segmentation on the Michigan-Columbia Dataset}
\label{section:nor_data}

\begin{table*}[bp]
    \centering
\ra{1.3}
\caption{Segmentation results on Michigan-Columbia Dataset. Standard deviation in parentheses, all values in \%. Results not gathered by us are taken from \cite{Oskal2019}}
\resizebox{\textwidth}{!} {%
\begin{tabular}{@{}rrrrcrrrcrrrrrrcrrrcrrrcrrrcrrrcrrrcrrrcrrrcrrrc@{}}\toprule
& PPV/Precision & \phantom{a} & Sensitivity/Recall & \phantom{a} & Dice-Score/F1-Score & \phantom{a} & Matthews Correlation & \phantom{a} & Mean-IoU & \phantom{a} & Accuracy &&\\ \midrule
Heggerty at al.~\cite{haggerty2014segmentation} & 35 $(\pm\  22)$ & & \textbf{99} $(\pm\  1)$ & & 47 $(\pm\  25)$ & & 52 $(\pm\ 22)$ & & - & & - \\
Lu et al.~\cite{lu2015automated} & 73 $(\pm\  27)$ & & 31 $(\pm\  31)$ & & 39 $(\pm\  33)$ & & 42 $(\pm\  31)$ & & - & & -  \\
Xu et al.~\cite{hongming2015epidermis} & 69 $(\pm\  20)$ & & 38 $(\pm\  32)$ & & 45 $(\pm\  28)$ & & 47 $(\pm\  27)$  && - & & -  \\
Kłeczek et al. \cite{kleczek2017automated} & 65 $(\pm\  25)$ & & 84 $(\pm\  26)$ & & 68 $(\pm\  23)$ & & - & & - & & -  \\
U-Net \cite{Oskal2019} & 89 $(\pm\  16)$ & & 92 $(\pm\  10)$ & & \textbf{89} $(\pm\  13)$ & & \textbf{89} $(\pm\  11)$ & & - & & -  \\
U-Net (Ours) & \textbf{92} $(\pm\  8)$ & & 80 $(\pm\  22)$ & & 83 $(\pm\  18)$ & & 83 $(\pm\  15)$ & & 86 $(\pm\  11)$ & & 99 ($\pm\  2)$  \\
EU-Net (Ours) & $91 (\pm 7)$& & $89 (\pm 20)$ & & $88(\pm 16) $& & $88 (\pm 14)$ & & $90(\pm 10)$ & & $99(\pm\ 1)$ \\
\bottomrule
\end{tabular}%
}
\label{nor_data}
\end{table*}

\begin{figure}[!h]
    \centering
    \begin{subfigure}[b]{0.4\textwidth}
         \centering
         \includegraphics[width=0.7\textwidth]{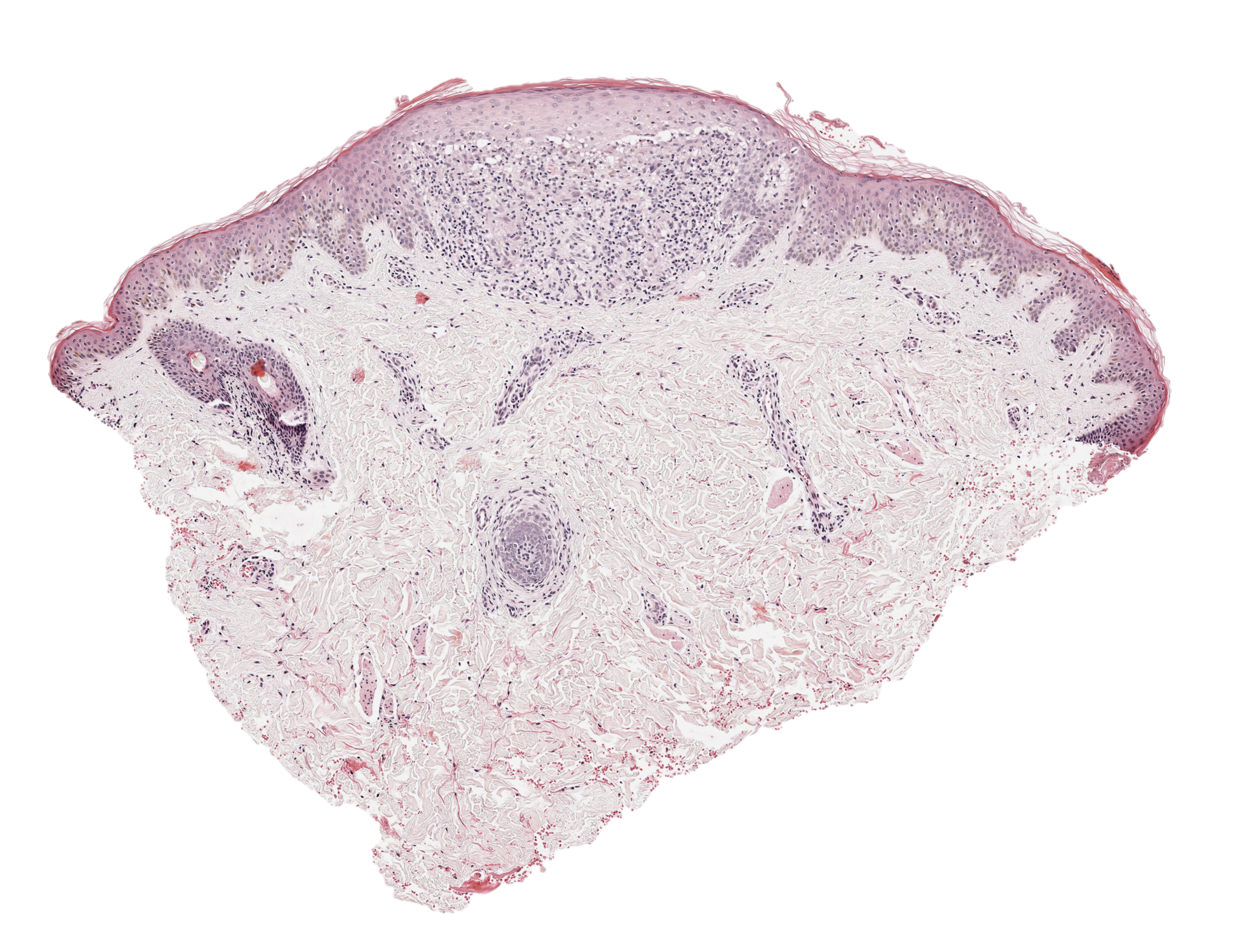}
         \caption{Example input WSI from the Michigan-Columbia dataset.}
         \label{inputUBCCase86}
     \end{subfigure}
     \hfill
     \begin{subfigure}[b]{0.4\textwidth}
         \centering
         \includegraphics[width=\textwidth]{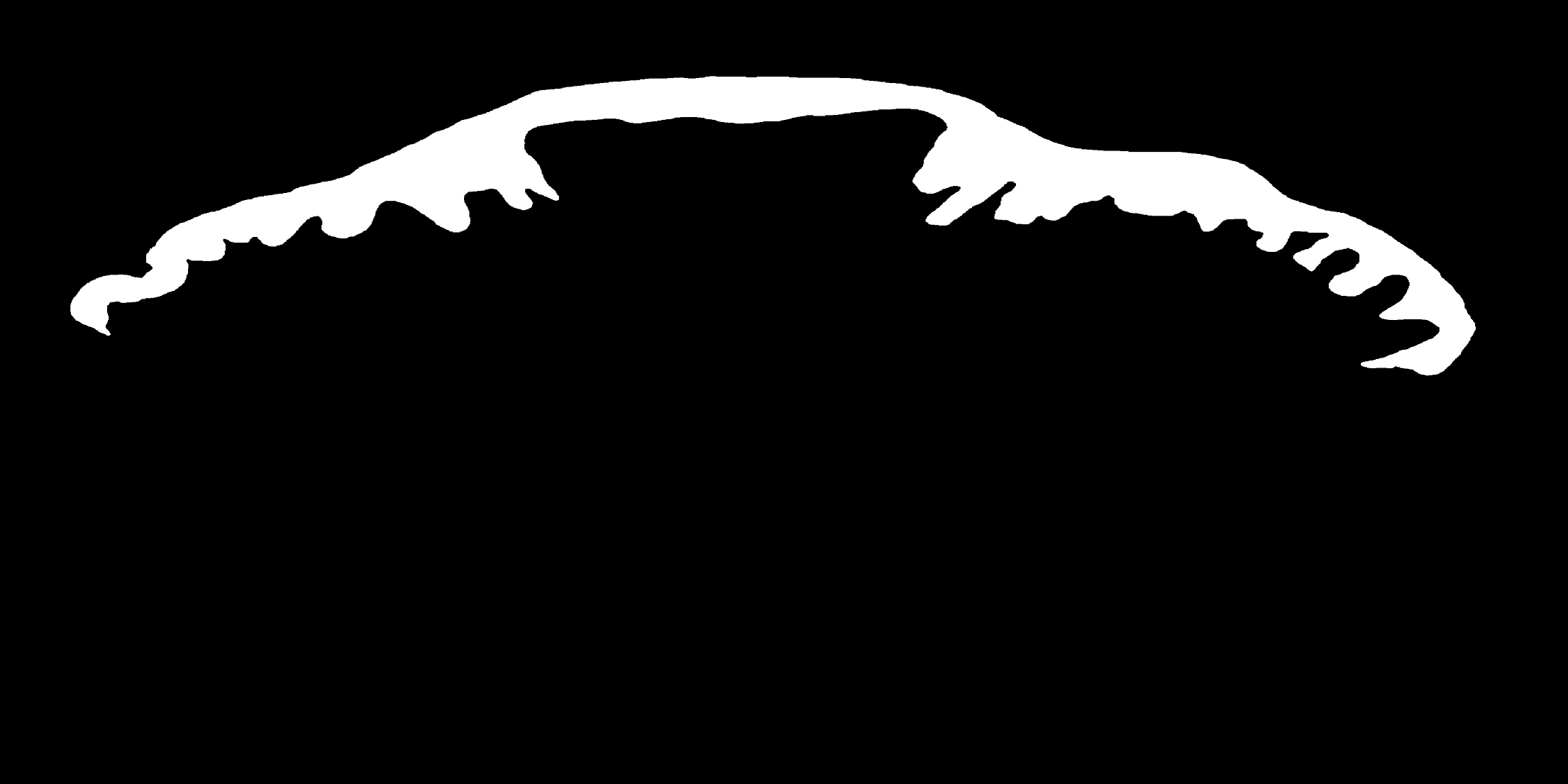}
         \caption{The corresponding label.}
         \label{labelUBCCase86}
     \end{subfigure}
     \hfill
     \begin{subfigure}[b]{0.4\textwidth}
         \centering
         \includegraphics[width=\textwidth]{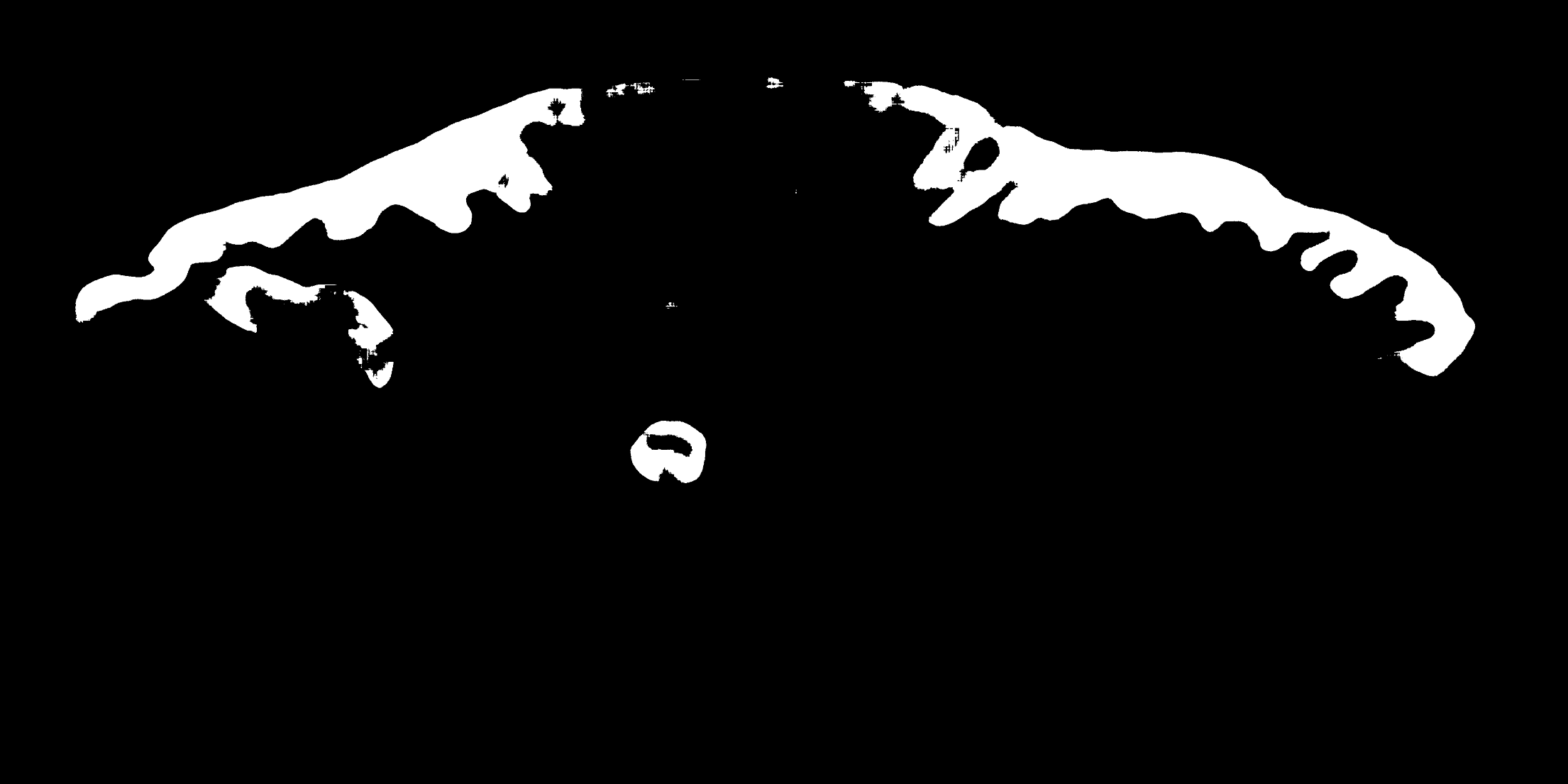}
         \caption{Prediction by our U-Net.}
         \label{unet_predictionUBCCase86}
     \end{subfigure}
     \hfill
     \begin{subfigure}[b]{0.4\textwidth}
         \centering
         \includegraphics[width=\textwidth]{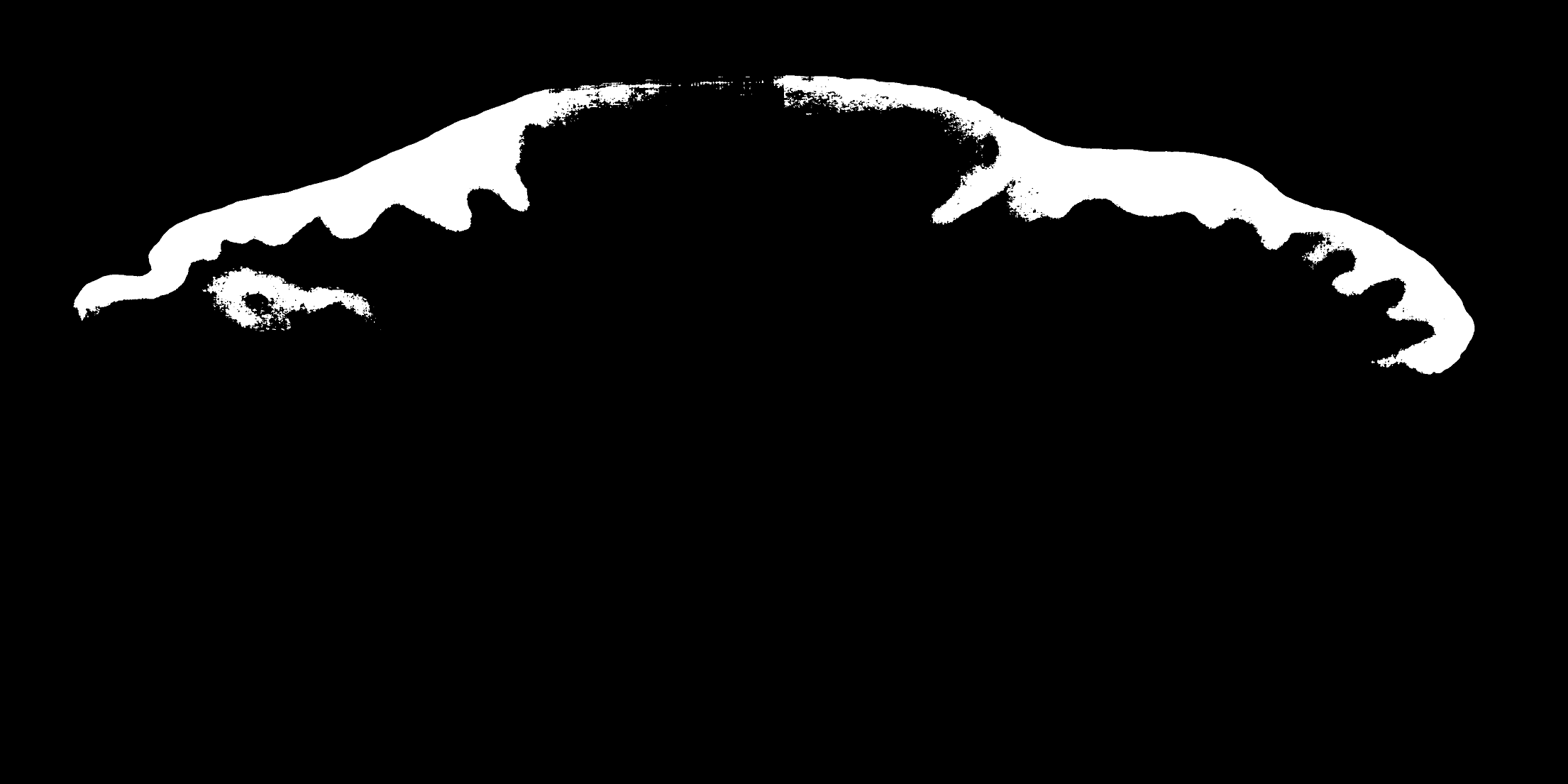}
         \caption{Prediction by our EU-Net.}
         \label{predictionUBCCase86}
     \end{subfigure}
     \caption{The figure shows an input WSI from the Michigan-Columbia dataset, its label (white represents epidermis and black represents ``rest'') and the predicted segmentation maps from our U-Net and EU-Net. This prediction resulted in the following metrics for the U-Net: Precision: $86\%$, Recall: $81\%$, F1-Score: $83\%$,  Matthews Correlation: $82\%$, Mean-IoU: $85\%$, Accuracy: $98\%$. And for the EU-Net: Precision: $91\%$, Recall: $79\%$, F1-Score: $85\%$,  Matthews Correlation: $84\%$, Mean-IoU: $86\%$, Accuracy: $98\%$.}
\label{fig:michigan_columbia_segmentation_predictions}
\end{figure}

As mentioned in Section~\ref{michigancolumbiadataset}, we use the same test set as \cite{Oskal2019} and preprocess the WSI in a similar way. Specifically, we take each WSI and resize its width and height to the nearest multiple of $4096$, then slice them into many patches of size $4096 \times 4096$. We make predictions on the patches, after which we join all the predictions back to the previous image size. Lastly we resize the images and predictions to half the width and half the height, before we calculate the metrics on them. Table~\ref{nor_data} shows the results of all the metrics we used to evaluate our models (see Appendix \ref{Appendix:metrics} for an explanation of the metrics). Note that PPV and Sensitivity do not constitute a good performance measure on their own as they are highly sensitive to class imbalance. They should only be taken into account as a pair. In Figure~\ref{fig:michigan_columbia_segmentation_predictions} we show an example of segmentation predictions of our U-Net and EU-Net on this dataset.

We can see that our EU-Net model is competitive with the adapted U-Net of \cite{Oskal2019}. It also clearly outperforms our standard implementation of a U-Net \cite{DBLP:journals/corr/RonnebergerFB15}. Moreover, our models outperform all other previous approaches with classical computer vision. The discrepancy in results between both U-Nets is due to the fact that we used an almost unaltered U-net from \cite{DBLP:journals/corr/RonnebergerFB15} while \cite{Oskal2019} mention they adapted the original U-Net and also used post-processing methods. In addition, due to the random nature of the patch extraction technique, we may have had a slightly different training set.

Given that our models are able to perform competitively  on a well established task, this will provide some crucial insights into the difficulty of performing semantic segmentation on the MF/E-Segmentation dataset.

\subsection{Segmentation on the MF/E-Segmentation Dataset} 
\begin{table*}[tp]\centering
\ra{1.3}
\caption{Segmentation results on the MF/E-Segmentation Dataset. Standard deviation in parentheses, all values in \%.}
\resizebox{\textwidth}{!} {%
\begin{tabular}{@{}rrrrcrrrcrrr@{}}\toprule
\cmidrule{1-5}
& \phantom{abcdefghijklmnopqrstuvwxyz} & Matthews Correlation & \phantom{abcdefghij} & Mean-IoU & \phantom{abcdefghij} & Accuracy &&\\ \midrule
U-Net (Ours)& & 70 $(\pm\ 11)$ & & 58 $(\pm\ 6)$ & &  98 $(\pm\ 1)$\\ 
EU-Net (Ours) & & \textbf{83} $(\pm\ 9)$ & & \textbf{69} $(\pm\ 9) $ & & \textbf{99} $(\pm\ 1)$ &&\\
\bottomrule
\end{tabular}%
}
\label{our_data}
\end{table*}

\begin{figure}[!h]
    \centering
    \begin{subfigure}[b]{0.4\textwidth}
         \centering
         \includegraphics[width=.6\textwidth]{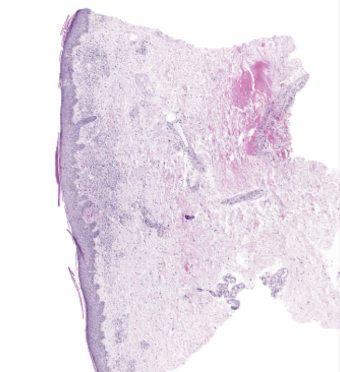}
         \caption{Example input WSI from the MF/E-Segmentation dataset.}
         \label{pfalz_inputE9_slide6}
     \end{subfigure}
     \hfill
     \begin{subfigure}[b]{0.4\textwidth}
         \centering
         \includegraphics[width=\textwidth]{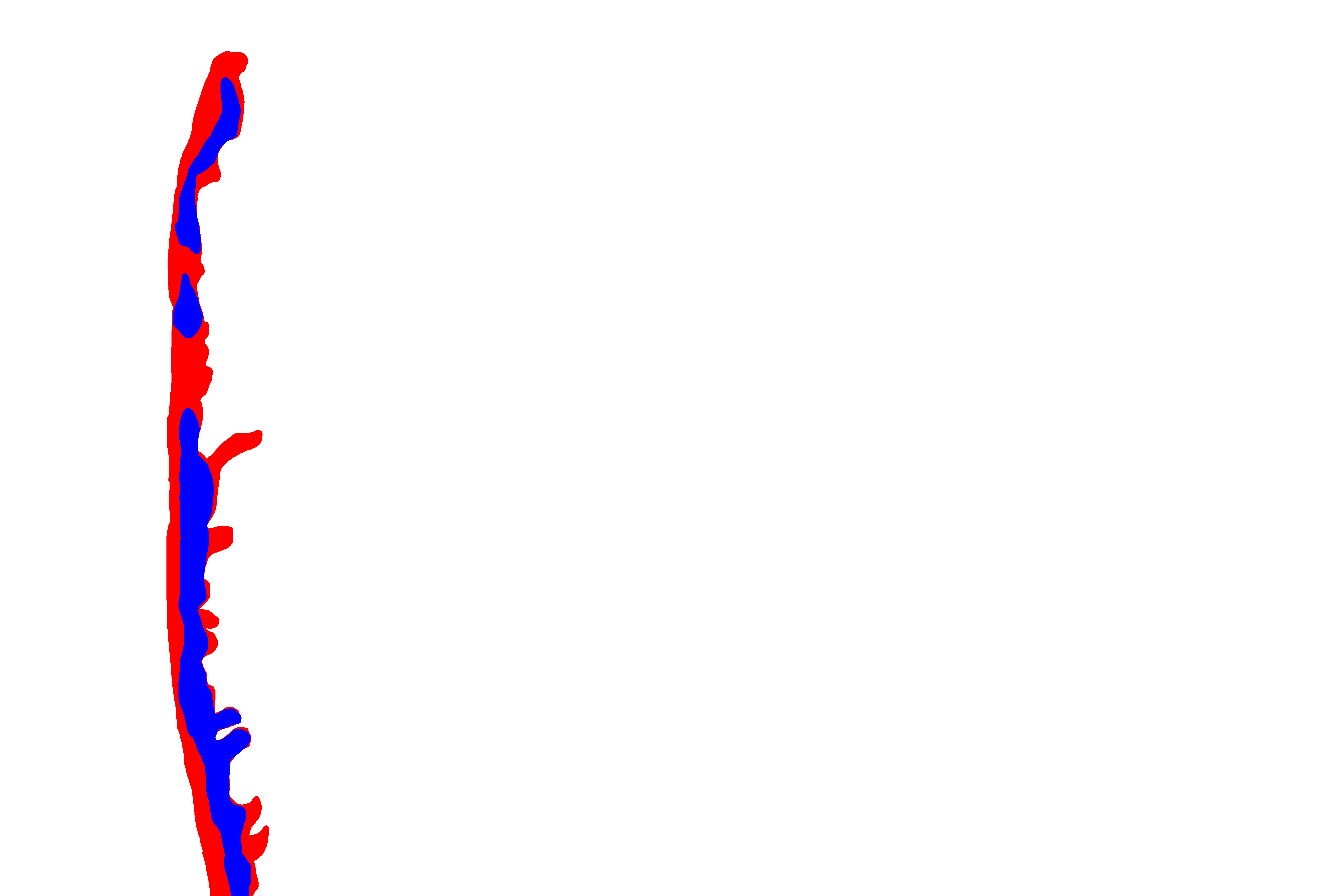}
         \caption{The corresponding label.}
         \label{pfalz_labelE9}
     \end{subfigure}
     \hfill
     \begin{subfigure}[b]{0.4\textwidth}
         \centering
         \includegraphics[width=\textwidth]{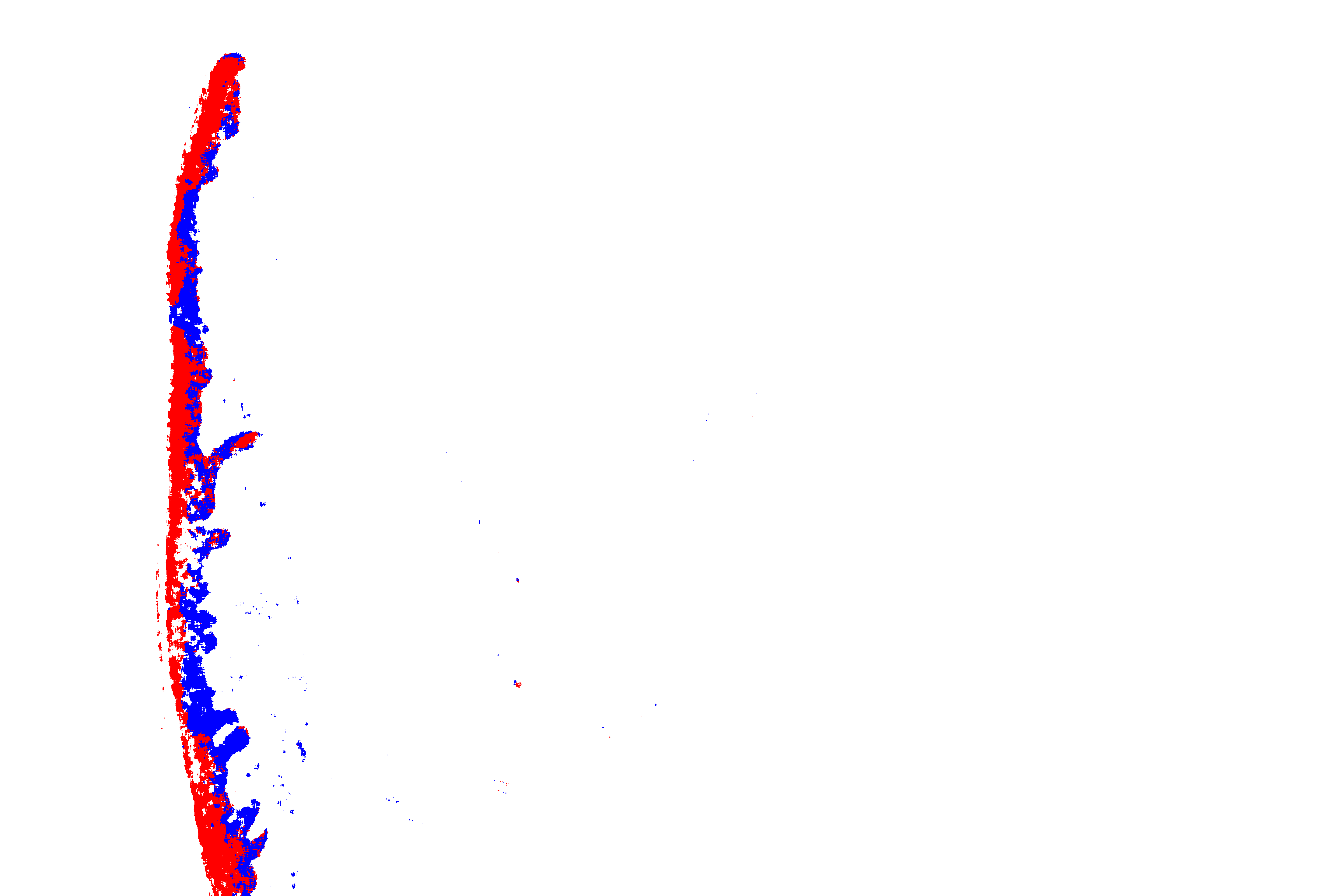}
         \caption{Prediction by our U-Net.}
         \label{unet_pfalz_predictionE9}
     \end{subfigure}
     \hfill
     \begin{subfigure}[b]{0.4\textwidth}
         \centering
         \includegraphics[width=\textwidth]{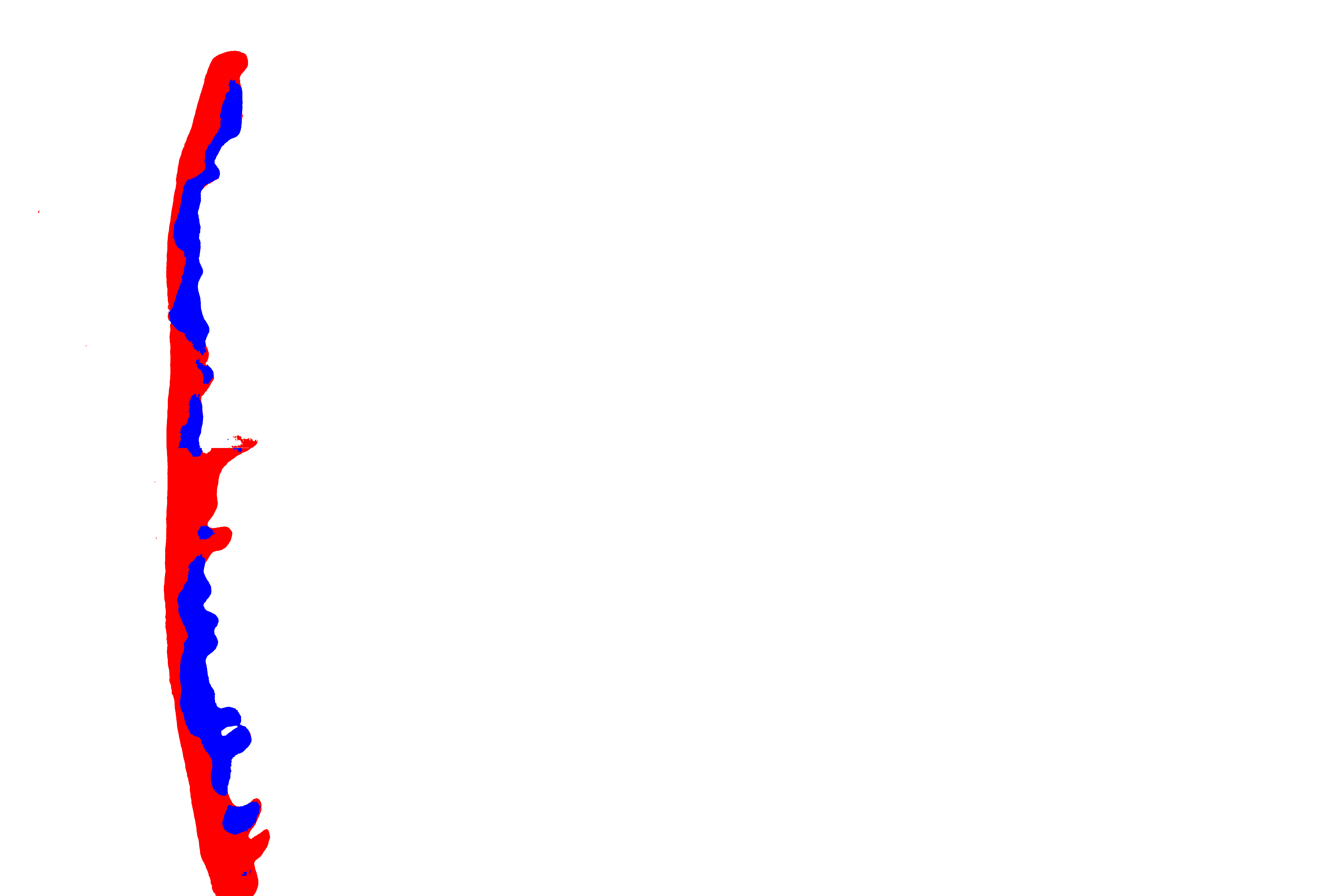}
         \caption{Prediction by EU-Net.}
         \label{pfalz_predictionE9}
     \end{subfigure}
     \caption{We can see an input WSI from the MF/E-Segmentation dataset, its label (blue represents spongiosis, red represents epidermis and white represents ``rest''). and the predicted segmentation maps from our U-Net and EU-Net. This prediction resulted in the following metrics for the U-Net:  Matthews Correlation: $74\%$, Mean-IoU: $63\%$, Accuracy: $99\%$. And for the EU-Net:  Matthews Correlation: $82\%$, Mean-IoU: $69\%$, Accuracy: $99\%$.}
\label{fig:mf_e_dataset_segmentation_predictions}
\end{figure}

Table~\ref{our_data} shows the results of our two models on the MF/E-Segmentation dataset and Figure~\ref{fig:mf_e_dataset_segmentation_predictions} shows example segmentation predictions of our U-Net and EU-Net. We see again that the EU-Net outperforms the U-Net significantly. We have no previous state of the art to compare them with, but given the results shown in Section~\ref{section:nor_data}, it is reasonable to assume that this is a much more challenging segmentation scenario. This might be due to the fact that this is a 3-class segmentation task, with spongiosis being contained in the epidermis class.

\subsection{Classification on the MF/E-Classification Dataset} \label{results:classification}

\begin{table*}[bp]\centering
\ra{1.2}
\caption{Classification on the MF/E-Classification Dataset. All values in \%}
\resizebox{\textwidth}{!} {%
\begin{tabular}{@{}rrrcrrrcrrr@{}}\toprule
& \phantom{abcdefghijkl} & Accuracy & \phantom{abc} & Precision/PPV & \phantom{abc} & Recall/Sensitivity & \phantom{abc} & F1 Score &&\\ \midrule
Baseline Eczema & & 68 & & - & & - & & - &&\\
\cmidrule{3-9}
\textbf{Model-Binary Cross Entropy} \\ w/o Segmentation Map& & 65 $(\pm\ 7)$ & & 54 $(\pm\ 33)$ & & \textbf{17} $(\pm\ 8)$ & & 22 $(\pm\ 9)$ &&\\
with Segmentation Map & & \textbf{71} $(\pm\ 3)$ & & \textbf{75} $(\pm\ 27)$ & & \textbf{17} $(\pm\ 3)$ & & \textbf{27} $(\pm\ 4)$ &&\\
\cmidrule{1-9}
Baseline MF & & 32 & & - & & - & & - &&\\
\cmidrule{3-9}
\textbf{Model-Cosine Similarity} \\ w/o Segmentation Map& & 35 $(\pm\ 3)$ & & \textbf{32} $(\pm\ 1)$ & & \textbf{91} $(\pm\ 13)$ & & \textbf{47} $(\pm\ 3)$ &&\\
with Segmentation Map & & \textbf{38} $(\pm\ 7)$ & & \textbf{32} $(\pm\ 2)$ & & 88 $(\pm\ 10)$ & & \textbf{47} $(\pm\ 3)$ &&\\
\bottomrule
\end{tabular}%
}
\label{table:classification_data}
\end{table*}

Due to the very limited data, we opted for using 5 fold cross-validation for a better picture of the model’s performance instead of using a very small test set. The same precautions regarding patient split and imbalanced data, as described in section~\ref{section:Seg_data}, were taken. No hyper-parameter tuning was performed, as this would bias the results. The results can be found in Table~\ref{table:classification_data}. Note that for the calculations of the metrics we took MF to be the positive class.

Using two different loss functions (binary cross entropy \& cosine similarity \cite{barz2019deep}) with the same architecture, we were able to achieve two different local minima, each with its own advantages. We first investigate, for both of the models, whether concatenating the segmentation map to the WSI input improves the classification performance. We will then take a closer look at how the two models compare.

\subsubsection{Classification with Segmentation Map}

We trained both models once with only the WSI as input and once with the segmentation map concatenated to it. For \textit{Model-Binary Cross Entropy}, using the additional segmentation map as input substantially improves all metrics except for recall where the performance remains unchanged. The \textit{Model-Cosine Similiarity} benefits from adding the segmentation map in terms of accuracy but suffers a comparable decrease in recall. The results suggest that using the segmentation map as an additional input can help classification performance.
For the remainder of the discussion we will refer only to the models using the additional segmentation map as input.

\subsubsection{Model Comparison}

\textit{Model-Binary Cross Entropy} predicts almost always Eczema, but when it predicts MF, it is highly accurate (high precision). \textit{Model-Cosine Similarity}, on the other hand, is very good at identifying all the MF cases (high recall). This is illustrated in Table~\ref{table:classification_data}, where each of the models is compared to its respective baseline, which is a dummy classifier which will always predict the same class. Both models have a huge difficulty capturing differences in the classification, as can be seen by the accuracy, which in both cases is barely above their respective baselines.

We believe that such models can still be useful. \textit{Model-Cosine Similarity}, for example, could be used in a pathology lab where MF vs. Eczema distinctions need to be made. It could filter out the obvious negative (Eczema) cases, leading to a lightened workload on pathologists and freeing their time for more pressing cases. The \textit{Model-Binary Cross Entropy}, on the other hand, could be used in routine checks where normally no pathologist would be available. Not many false positives would be generated by the model, but it will catch some MF cases that would otherwise have been overlooked, thereby potentially saving lives.

We also want to reiterate the misclassification rate of 21.51$\%$ \cite{PIMPINELLI20051053} of pathologists on a different dataset, which highlights the difficulty of this task, even for specialists. Nevertheless, we acknowledge that our results still leave much room for improvement. However, it seems difficult to improve resorting only to fine-tuning our models. We believe that having a larger dataset would definitely increase the chances of accurately classifying between MF and Eczema.

\section{Discussion}
We are able to show that our EU-Net architecture performs competitively on the dataset of \cite{Oskal2019} in the relevant metrics. We then use this model on the MF/E-Segmentation dataset which contains annotations for epidermis, spongiosis and ``rest''. The lower scores on the  MF/E-Segmentation dataset, in comparison with the Michigan-Columbia dataset, suggest that this segmentation task is significantly harder. Nevertheless, when looking at actual segmentation maps (see Figure~\ref{fig:mf_e_dataset_segmentation_predictions}), one can still imagine these predictions being helpful to a pathologist. 

The binary classification results are not yet at a level where pathologists could use them to make predictions. This was expected given the high inter-rater variability of $48\%$ among pathologists. However, as mentioned before in section~\ref{results:classification}, we believe that the combination of both models \textit{Model-Binary Cross Entropy} and \textit{Model-Cosine Similarity} could be used together to provide value to pathologists.

We believe that our models are not only valuable diagnostic tools, but that they provide interpretable results. The resulting segmentation map generated will aid pathologists in discerning if the models' output is reasonable and if the resulting prediction is trustworthy. This will add a level of confidence which is not granted in many diagnostic tools, facilitating its adoption.

\section{Future Work}
Many of the difficulties we faced were due to the limited nature of our data set. The continued effort of labeling WSI slides should definitely result in better semantic segmentation of the critical areas. In addition, there are a few key ideas that we did not yet explore. First, some post-processing techniques could be applied to our segmentation predictions. Second, the concatenation of multiple data sets, even if they lack certain key labels, could provide a valuable data augmentation to such a data starved task. One could again use a transfer-learning approach and pre-train on the data of \cite{Oskal2019} before then training on our own data, and vise versa. For the classification task, one could try to combine the presented models in order to leverage their individual strengths.

\subsubsection{Acknowledgements}
We would like to thank everybody at the Kempf und Pfaltz Histologische
Diagnostik lab that provided crucial insights about the
task, all the necessary data and for their time and commitment to this
collaboration. Finally, we would like to thank
PAIGE\footnote{\url{paige.ai}} for providing the software that was
used to annotate the WSI.

\bibliographystyle{splncs04}
\bibliography{references}
\newpage
\appendix

\section{Appendix}
\subsection{Metrics}\label{Appendix:metrics}
Here we provide the definitions of the metrics used in evaluating our models. In the multi-class scenario, the definitions of these metrics alter slightly. For $n$ classes, with $TP_n,\ n \in \{1...n\}$ representing the True Positives for each class, we define $TP = \sum_{1}^{n}{TP_n}$. Similarly, we define $TN = \sum_{1}^{n}{TN_n}$ for True Negatives, $FP = \sum_{1}^{n}{FP_n}$ for False Positives and $FN = \sum_{1}^{n}{FN_n}$ for False Negatives.

\begin{equation}
PPV/Precision = \mathcal{A}_{PPV} = \frac{TP}{TP + FP} \\
\end{equation}

\begin{equation}
Sensitivity/Recall = \mathcal{A}_{SEN} = \frac{TP}{TP + FN} \\
\end{equation}

\begin{equation}
\text{\textit{Dice-Score/F1-Score}} = 2 \cdot \frac{\mathcal{A}_{PPV} \cdot \mathcal{A}_{SEN}}{\mathcal{A}_{PPV} + \mathcal{A}_{SEN}}\\
\end{equation}

\begin{equation}
Matthews Correlation = \frac{TP \cdot TN - FP \cdot FN}{\sqrt{(TP+FP)(TP+FN)(TN+FP)(TN+FN)}}\\
\end{equation}

\begin{equation}
\text{\textit{Mean-IoU}} = \frac{TP}{TP + FP + FN}\\
\end{equation}

\begin{equation}
Accuracy = \frac{TP + TN}{TP + TN + FP + FN}\\
\end{equation}
The intuition behind using multiple metrics is that none of them individually provides the full picture. Accuracy is the percentage of correctly labeled pixels. It is a simple measure but that can be misleading in settings with high class imbalance as we have here. Precision is the fraction of actual true instances among all predicted true instances. Sensitivity is the fraction of true instances that were predicted as true. It's easy to see how these metrics complement each other. The Dice-Score is a harmonic average over these two metrics, which allows for a better interpretation of the individual scores. The Mean-IoU complements this, providing a simple measure of area of intersection between prediction and label, divided by union of the areas.  Lastly, Matthews Correlation Coefficient is a balanced measure and can be used in high class imbalance cases.

\pagebreak

\subsection{EfficientNet Training Parameters} \label{Appendix:efficientnet}
From the library provided by \cite{Yakubovskiy:2019}, we experimented with different encoders and general segmentation architectures such as Linknet \cite{linknet}, Pyramid Scene Parsing Network \cite{pyramidscenepasing} and Feature Pyramid Network \cite{featurepyramidnetwork}. Eventually, the results of experiments indicated that the combination of EfficientNet-B7 and U-Net was best suited for our task.

On the dataset of \cite{Oskal2019} the model was trained for 24 epochs. On the MF/E-Segmentation dataset it was trained for 6 epochs. The rest of the configurations were the same for both datasets with a batch size of 4 and the Adam optimizer \cite{kingma2014adam}. We used a learning rate of 0.001 with a decay rate of 0.96 every 50000 steps. We additionally used image augmentation which randomly rotated the images by multiples of 90 degrees and then with a probability of $0.5$ flipped the image top-down and with the same probability also left-right.

\subsection{U-Net Architecture} \label{Appendix:unet}

\begin{figure}[!hb]
    \centering
    \includegraphics[width=0.6\textwidth]{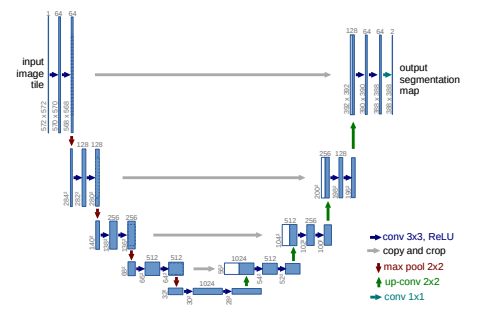}
    \caption{``U-net architecture (example for 32x32 pixels in the lowest resolution). Each blue box corresponds to a multi-channel feature map. The number of channels is denoted on top of the box. The x-y-size is provided at the lower left edge of the box. White boxes represent copied feature maps. The arrows denote the different operations.'' Caption and Figure from \cite{DBLP:journals/corr/RonnebergerFB15}}
    \label{fig:unet}
\end{figure}

\pagebreak

\subsection{EU-Net Architecture} \label{Appendix:eunet}
\begin{figure}[!hb]
    \centering
    \includegraphics[width=.5\textwidth]{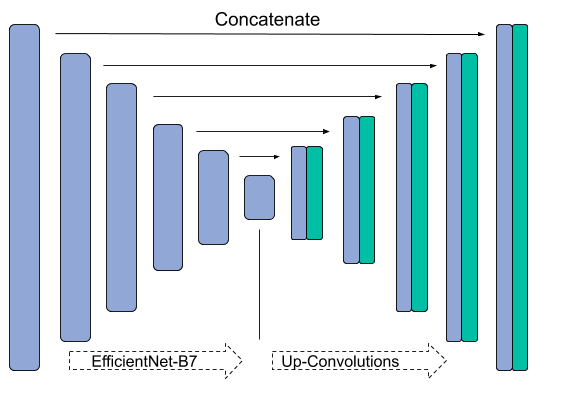}
    \caption{EU-Net Architecture. Note that the encoder is a symbolic representation of the EfficientNet-B7, as the actual model is much larger. }
    \label{fig:eunet_arch}
\end{figure}

\end{document}